\documentclass[preprint,nofootinbib,superscriptaddress]{revtex4}
\usepackage{amsmath,amsthm,amssymb}
\usepackage[dvips]{graphicx}
\renewcommand{\slash}{\displaystyle{\not}}
 
\begin{document}
\preprint{
 {\vbox{
 \hbox{MADPH--05--1443}
 \hbox{hep-ph/0510257}}}}
 
\title{\quad \\
Neutralino annihilation beyond leading order}

\author{Vernon~Barger}
\email{barger@pheno.physics.wisc.edu}

\affiliation{Department of Physics, University of Wisconsin, 1150 University
Avenue, Madison, Wisconsin 53706 USA}

\author{Wai-Yee~Keung}
%% \email{keung@tigger.cc.uic.edu}

\affiliation{Physics Department, University of Illinois at Chicago, 
Illinois 60607--7059 USA}

\author{Heather~E.~Logan}
\email{logan@physics.carleton.ca}

\affiliation{Department of Physics, University of Wisconsin, 1150 University
Avenue, Madison, Wisconsin 53706 USA}

\affiliation{Physics Department, Carleton University, 
Ottawa, Ontario K1S 5B6 Canada}
 
\author{Gabe~Shaughnessy}
\email{gshau@physics.wisc.edu}
 
\author{Adam~Tregre}
\email{tregre@pheno.physics.wisc.edu}
 
\affiliation{Department of Physics, University of Wisconsin, 1150 University
Avenue, Madison, Wisconsin 53706 USA}

\begin{abstract}
High-precision measurements of the relic dark matter
density and the calculation of dark matter annihilation branching fractions in the sun or the galactic halo today motivate the computation of the neutralino annihilation cross
section beyond leading order.
We consider neutralino annihilation via squark exchange and 
parameterize the effective annihilation vertex as a dimension-six 
operator suppressed by two powers of the squark mass and
related to the divergence of the axial vector current of the
final-state quarks.  Since the axial vector current is conserved at 
tree level in the limit of massless quarks, 
this dimension-six operator contains a suppression
by the quark mass.  The quark mass suppression can be lifted in two
ways: (1) by corrections to the dimension-six operator involving the
anomalous triangle diagram, and (2) by going to dimension-eight.
We address the first of these possibilities by evaluating the anomalous 
triangle diagram, which contributes
to neutralino annihilation to gluon pairs.  We relate the triangle diagram  
via the anomaly equation to the decay of a pseudoscalar into 
two gluons and use the Adler-Bardeen theorem to extract the
next-to-leading order (NLO) QCD corrections to $\chi \chi \to gg$
%neutralino annihilation into gluon pairs 
from the known corrections to pseudoscalar decay.
The strong dependence of the dominant $\chi \chi \to q \bar q$ cross 
section on the relative velocity of the neutralinos makes these NLO 
corrections unimportant at $\chi$ decoupling but significant today.
\end{abstract}

\date{November 18, 2005}
 
\maketitle
%--------------------------------------------------------------

\section{Introduction}

The presence of non-baryonic dark matter in the universe is compelling
evidence for physics beyond the Standard Model.  While many very different
models have been proposed to explain the dark matter 
(wimps, axions, wimpzillas,
modified gravity, etc.) \cite{dmrev}, the thermal production of 
stable weakly-interacting 
particles with weak-scale mass remains as the most attractive and predictive 
explanation for the observed dark matter relic abundance, and further allows
the solution to the dark matter problem to be linked to the solution to
the hierarchy problem and tested at current and future collider experiments.  Supersymmetry provides an especially attractive explanation with the lightest
supersymmetric particle (LSP) as the dark matter candidate.
Once such a weakly-interacting massive particle ($\chi$) 
has been discovered and its
couplings measured, it will be possible to compute its annihilation cross
section (which controls its thermal relic abundance) and compare to the
measured dark matter abundance to test our understanding of the microphysics 
of dark matter.   The cosmological dark matter abundance is already measured
at the 10\% level to be \cite{Bennett:2003bz} 
%%
%\begin{equation}
$$
\Omega_{\rm CDM}h^2 =0.12\pm0.01  \quad(\hbox{SDSS+WMAP}) \ 
$$
%\end{equation} 
%%
and future cosmic microwave background 
experiments such as PLANCK 
expect to improve this to the few-percent level \cite{Tauber}.  
In order to match the expected few-percent
precision of the cosmological measurements, we need both high-precision
inputs from the colliders and high-precision calculations of the neutralino
annihilation cross section.  The latter requirement means going beyond
leading order.  High-precision calculations of the relic abundance are thus 
needed to match the microphysics onto cosmology.

Another important physics application of these higher order QCD 
corrections is the calculation of signals from WIMP annihilations in 
the galatic halo or in the interior of the sun.  The NLO corrections 
are potentially important in the evaluation of the branching fractions 
for the observable gamma ray and neutrino signals, respectively.  As we 
shall see, these NLO corrections turn out to be much more important
for the calculation of the observable gamma ray and neutrino signals
than in the calculation of the relic density because of the strong 
dependence of the tree-level annihilation cross-section on the relative 
velocity of the neutralinos.

\subsection{Neutralino annihilation cross section}

\begin{figure}
\resizebox{\textwidth}{!}{
\includegraphics{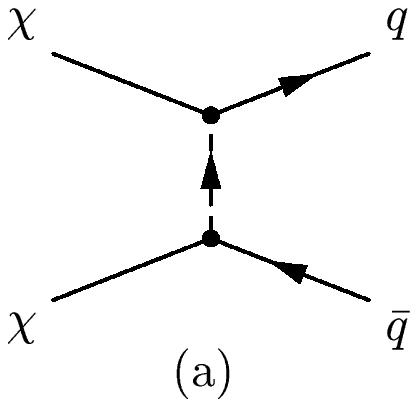}
\includegraphics*{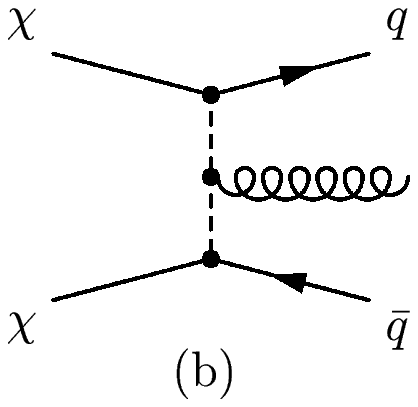}
\includegraphics*{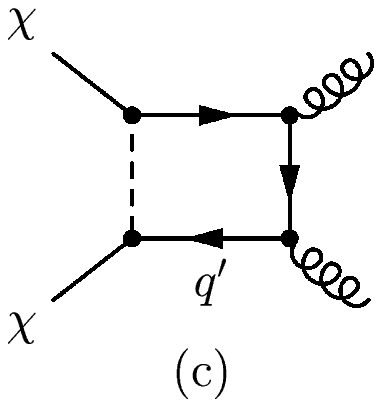}
\includegraphics*{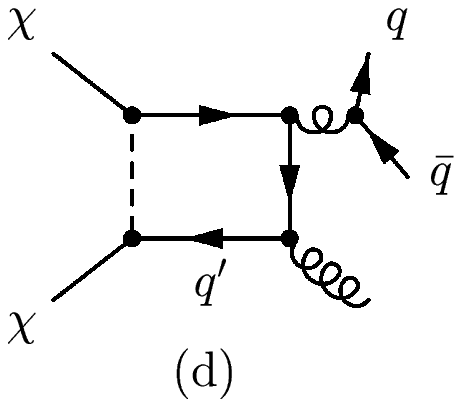}
\includegraphics*{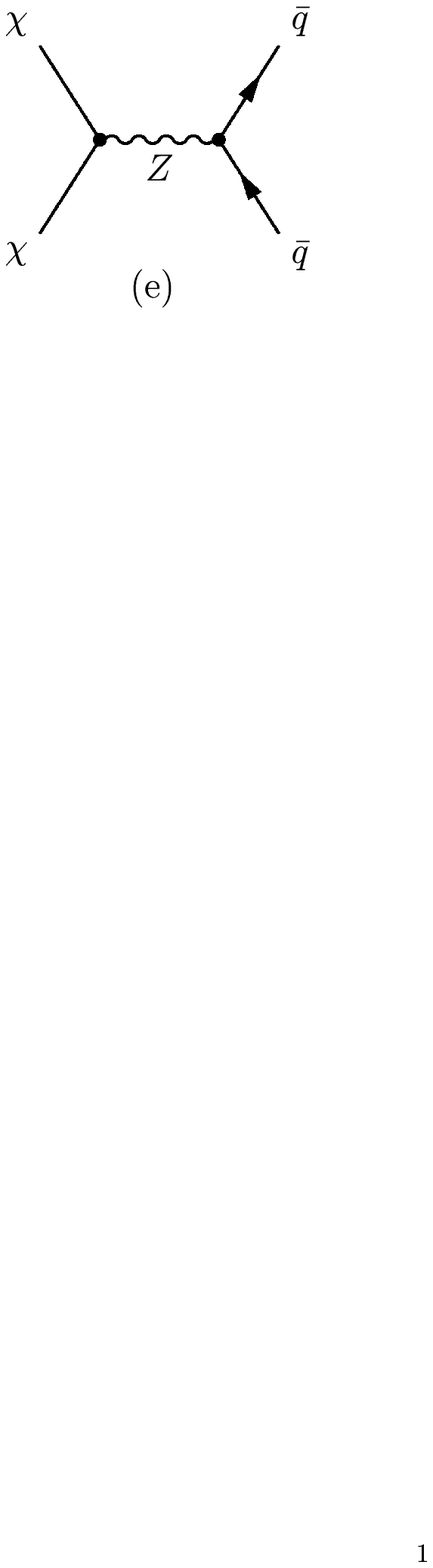}
}
\resizebox{\textwidth}{!}{
\includegraphics*{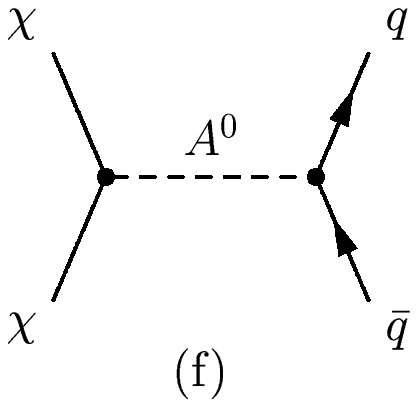}
\includegraphics*{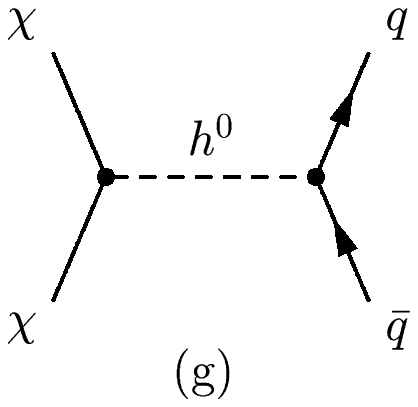}
\includegraphics*{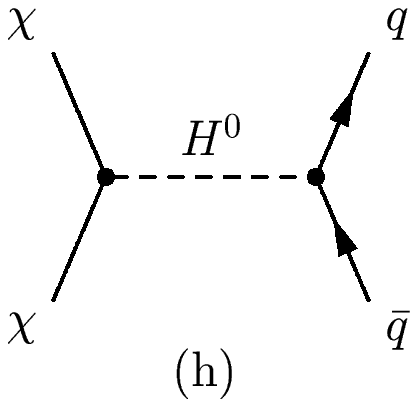}
\includegraphics*{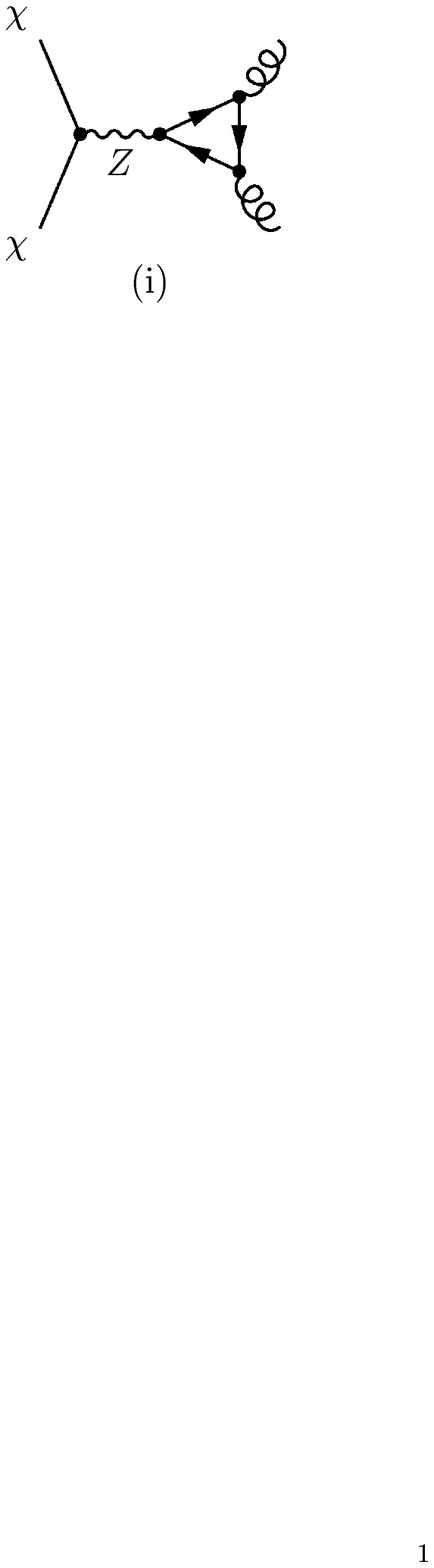}
\includegraphics*{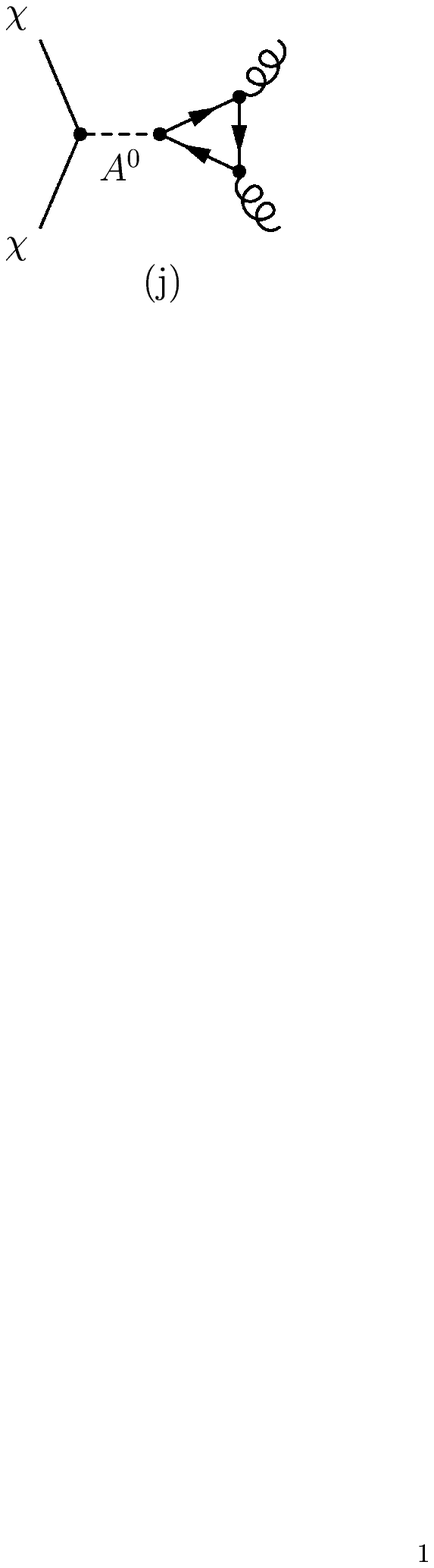}
}
\caption{Feynman diagrams that contribute to the total annihilation cross 
section: (a) the tree level diagram, (b-d) $t$-channel squark exchange, 
and (e-j) $s$-channel $Z$ and Higgs exchanges. }
\label{fig:fd}
\end{figure}

The behavior of the annihilation cross section depends on the composition 
of the neutralino.  Throughout this paper we assume that the LSP is largely 
gaugino as motivated by mSUGRA models \cite{ref:sps}.  The processes that 
contribute to the cross section up to order  $\alpha_s^2$ and one loop are 
shown in Fig.~\ref{fig:fd}.  The tree-level diagram is shown in 
Fig.~\ref{fig:fd}(a).  
Figures~\ref{fig:fd}(b-d) show the diagrams with $t$-channel squark 
exchange, whereas (e-j) show the diagrams with $s$-channel $Z,H^0,h^0,A^0$ 
exchanges.  The gauge and Higgs bosons couple to the Higgsino part of the LSP 
and thus their contributions are suppressed for a mostly-gaugino 
neutralino.\footnote{The Higgsino fraction suppression can be removed 
at the cost of going to one loop \cite{Djouadi:2001kb}.}
The corresponding suppression 
factors for the $s$- and $p$-wave terms in the cross section are given in 
Table~\ref{tab:dependence}.

\begin{table}
\begin{tabular}{cccc}
\hline\hline
Diagram & $\chi$ mixing & $s$-wave & $p$-wave \\
\hline
(a) & $\left[ \text{Gaugino}\right]^4$ 
    & $[ m_q/\widetilde M^2 ]^2$ 
     & $v^2 [ m_{\chi}/\widetilde M^2 ]^2$ \\
(b) & $\left[ \text{Gaugino}\right]^4$ 
    & $\alpha_s [ m_q/\widetilde M^2 ]^2 
       + \alpha_s [m_{\chi}^3/\widetilde M^4 ]^2$ & \\
(c) & $\left[ \text{Gaugino}\right]^4$   & $\alpha_s^2 [ m_{\chi}/\widetilde M^2 ]^2$ & \\
(d) & $\left[ \text{Gaugino}\right]^4$   & $\alpha_s^3 [ m_{\chi} /\widetilde M^2 ]^2$ & \\
(e) & $\left[ \text{Higgsino} \right]^4$
     & $[ m_q/(s-m_Z^2) ]^2$ & \\
(f) & $\left[ \text{Gaugino} \times \text{Higgsino} \right]^2$ 
     & $[(m_q/m_W) m_{\chi} /(s-m_A^2) ]^2$ & \\
(g) & $\left[ \text{Gaugino} \times \text{Higgsino} \right]^2$
     & 0 & $v^2 [ (m_q/m_W) m_{\chi}/(s-m_h^2) ]^2$ \\
(h)  & $\left[ \text{Gaugino} \times \text{Higgsino} \right]^2$
     & 0 & $v^2 [ (m_q/m_W) m_{\chi} /(s-m_H^2) ]^2$ \\
 (i)  & $\left[ \text{Higgsino}\right]^4$ 
     & $\alpha_s^2 [ m_{\chi}/m_Z^2 ]^2$ 
     &  \\
 (j)  & $\left[ \text{Gaugino} \times \text{Higgsino} \right]^2$
     & $\alpha_s^2 [ m_{\chi}/(s-m_A^2) ]^2$ 
     &  \\
\hline\hline
\end{tabular}
\caption{Dependence of the cross section from each diagram 
on various suppression factors.  The diagrams are shown in Fig.~\ref{fig:fd}.
The $p$-wave contribution to the cross section is shown only when 
it is important
due to suppression or absence of the $s$-wave component.  
The column labeled ``$\chi$ mixing'' sketches the dependence of 
the cross section on the neutralino
composition.  Interference terms between the 
various diagrams carry a combination of the suppression factors 
corresponding to each diagram, and are not shown.  
We note that the $Z$-pole structure of diagram (i) is canceled by a 
numerator factor supplied in accordance with Yang's 
theorem \cite{Yang:1950rg,Rudaz:1989ij}.}
\label{tab:dependence}
\end{table}

\subsection{The anomaly equation}
\label{sec:anomalyequation}

The leading contribution to neutralino annihilation via exchange of
a squark of mass $\widetilde M$, shown in Fig.~\ref{fig:fd}(a),
can be reduced to an effective vertex described by a dimension-six operator
suppressed by $\widetilde M^2$,
\begin{equation}
  {\cal L} = (c/\widetilde M^2) {\cal O}_6 \ ,\quad
  {\cal O}_6=(\overline\chi\gamma_\mu\gamma_5 \chi)  
           (\overline q  \gamma^\mu\gamma_5  q) \ ,
\end{equation}
where $c$ is a dimensionless coefficient.  This dimension-six operator
corresponds to taking the leading term in the expansion of the squark
propagator in powers of $1/\widetilde M^2$; in particular, we 
work in the limit $m_{\chi}^2 \ll \widetilde M^2$.

In the static limit, where the relative velocity of the two
neutralinos can be neglected, 
the operator ${\cal O}_6$ is related to the divergence of the axial 
vector current of the quarks $\overline q q$: 
\begin{equation}
 {\cal O}_6 \to 
\left[ \overline\chi \ (i\gamma_5 / 2m_\chi)  \chi \right]  \  
 \left[\partial_\mu (\overline q \gamma^\mu\gamma_5  q) \right]\ .
 \end{equation}
In the massless quark limit, $m_q=0$, the axial vector current
is conserved at tree level,
$\partial_\mu (\overline q \gamma^\mu\gamma_5  q)=0 $,
and all tree amplitudes due to the dimension-six 
operator vanish; in particular,
radiating additional gluons cannot lift the suppression.
Even at the loop level,
for example diagrams involving the exchange of a virtual gluon,  the
suppression is still valid unless the anomalous triangle diagram is
involved. 
This is the well-known partially-conserved axial current (PCAC) condition.

Indeed, only through the anomalous loop diagrams is
the conservation of the axial vector current violated, in the form
\cite{ref:adlera}
\begin{equation}
  \partial_\mu (\overline q \gamma^\mu\gamma_5 q) = 
   2m_q \overline{q} i\gamma_5 q 
  + \frac{\alpha_s}{4\pi} G^{(a)}_{\mu\nu} \widetilde{G}^{(a)\mu\nu}
  \label{eq:anomaly}
\end{equation}
with 
$\frac{1}{2}\widetilde{G}_{\mu\nu} 
= \epsilon_{\mu\nu\alpha\beta} G^{\alpha\beta}$ 
denoting the dual color field strength tensor.  
The simplest such anomalous
diagram is shown in Fig.~\ref{fig:fd}(c).
Neglecting the mass of the internal quark $q^{\prime}$ and using
the anomaly equation, this diagram can be written in the form
%%%%%
\begin{equation}
  {\cal L}_{\rm eff} (\chi\chi\to gg) =
  \left(\frac{c/m_\chi}{2 \widetilde M^2}\right) 
  (\overline \chi \ i\gamma_5 \ \chi)
  \frac{\alpha_s}{4\pi}  G^{(a)}_{\mu\nu} \widetilde G^{(a)\mu\nu} \ ,
\end{equation}
for $m_{q^{\prime}} \ll m_\chi$.  In the opposite limit, 
$m_{q^{\prime}} \gg m_{\chi}$, the very heavy quark
decouples; the top quark contribution can be neglected if
$m_\chi \lesssim 100$ GeV.
The leading order (LO) calculation using the 
anomaly equation was first studied correctly by Ref.~\cite{Rudaz:1989ij} 
in the $\gamma\gamma$ channel in QED.

The gluonic decay amplitude of a fundamental pseudoscalar, $A^0 \to gg$,
is also related to the anomaly equation.  This decay proceeds through
a quark loop.
In the heavy quark limit, $m_Q \gg m_A$,
the divergence term on the left-hand side of Eq.~(\ref{eq:anomaly}) 
becomes insignificant, leading to,
\begin{equation}
  0 \simeq 2 m_Q \overline Q i \gamma_5 Q 
  + \frac{\alpha_s}{4 \pi} G^{(a)}_{\mu\nu} \widetilde{G}^{(a) \mu\nu}.
\label{eq:anom2}
\end{equation}
Note that because the Yukawa coupling of $A^0$ to the quark $Q$ is 
proportional to the quark mass, the $m_Q$ dependence of the $A^0 \to gg$ 
partial width drops out in the limit $m_Q \gg m_A$.

In contrast, the neutralino pair annihilates into two gluons via the 
anomaly diagram, which is dominated by light quarks, $m_q \ll m_{\chi}$.
In this limit, the term proportional to $m_q$ on the right-hand side
of Eq.~(\ref{eq:anomaly}) becomes insignificant, leading to,
\begin{equation}
  \partial_\mu \overline q \gamma^\mu\gamma_5 q \simeq
  0 + \frac{\alpha_s}{4\pi} G^{(a)}_{\mu\nu} \widetilde{G}^{(a)\mu\nu}.
\label{eq:anom3}
\end{equation}
Thus we see that two seemingly different processes, 
$A^0 \to gg$ through a heavy
quark loop and $\chi\chi \to gg$ through a light quark loop, are related
by the anomaly equation to the \emph{same} gluonic operator.
The Adler-Bardeen theorem \cite{ref:abj} guarantees that the 
anomaly equation, Eq.~(\ref{eq:anomaly}), is valid to all orders of $\alpha_s$.
One can take advantage of this anomaly property to obtain the higher-order
QCD corrections to $\chi\chi \to gg$ from the known results for
$A^0 \to gg$ at next-to-leading order (NLO)
from Ref.~\cite{Spira:1995rr}.
Note that, because of the non-abelian nature of the gauge field, the above
gluonic operator also incorporates tri-gluon amplitudes beyond leading order.

\subsection{Beyond dimension-six}

If we include higher terms in the $1/\widetilde M^2$ expansion, the
PCAC constraint will be lifted. The dimension-eight operator 
corresponding to an amplitude proportional to $1/\widetilde M^4$ survives 
even in the massless quark limit, $m_q\to 0$.
One can use this dimension-eight amplitude to 
calculate the rate of $\chi\chi\to q\bar q g$ from diagrams such
as Fig.~\ref{fig:fd}(b); a full calculation was done in 
Ref.~\cite{Drees:1993bh}.
However, the contribution to $\chi\chi \to q \bar q g$ from the dimension-six
operator due to the anomaly with a virtual gluon turning into a
quark pair [Fig.~\ref{fig:fd}(c)] suffers less $\widetilde M$ 
suppression and will dominate the dimension-eight term for 
$\widetilde M \gg m_\chi$ even though the order in $\alpha_s$ is higher.

The effective vertex of the dimension-eight operator for
$\chi\chi \to q\bar q g$ is
%%%
\begin{equation}
  {\cal L} = (c_8 g_s^2/\widetilde M^4) {\cal O}_8 \ , \quad
  {\cal O}_8=\epsilon_{\mu\nu\alpha\beta}G^{(c) \alpha\beta}
  (\bar q\gamma^\nu\gamma_5 T^{c} q) (\bar \chi \gamma^\mu\gamma_5 \chi),
  \label{eq:dim8}
\end{equation}
where $G^{(c) \alpha\beta}$ is the gluonic field strength for 
color index $c$ and $T^{c}$ is the corresponding SU(3) generator.
This operator has exactly the same form as that in Eq.~(6) 
of Ref.~\cite{Flores:1989ru} for the $\chi \chi \to f \bar f \gamma$ 
amplitude computed through explicit 
expansion of the propagators to order $1/\widetilde M^4$ in the limit 
$m_q=0$.

It is interesting to note that the amplitude for $\chi \chi \to q \bar q g$ 
from the dimension-six operator with one gluon splitting into $q \bar q$
[shown in Fig.~\ref{fig:fd}(d)] yields an operator of the same form as
in Eq.~({\ref{eq:dim8}).  This allows the interference term 
between this diagram and the dimension-eight process to be easily obtained.

\section{Calculation}

\subsection{The anomaly at leading order}

Since neutralinos are Majorana in nature, the initial state behaves as a 
pseudoscalar in the zero-velocity limit \cite{Bergstrom:1988}.  
In particular, for $v_{\rm rel} = 0$ the antisymmetrized 
neutralino spinors reduce to the projection operators \cite{Kuhn:1979}
\begin{equation}
	u(p_1)\bar{v}(p_2)-u(p_2)\bar{v}(p_1) = 
	( m_{\chi} + \displaystyle{\not}P) \gamma_5 
	=  m_{\chi} ( 1 + \gamma^0) \gamma_5.
	\label{eq:antisym}
\end{equation}
where $p_1$ and $p_2$ are the four-momenta of the incoming neutralinos
and $2P=p_1+p_2$.

We work in the limit of zero fermion mass, in which case the off-diagonal
terms in the squark mass matrices vanish %[see Eq.~(\ref{eq:squarkmass})] 
and the squark mass eigenstates coincide with the electroweak 
eigenstates $\widetilde q_L$ and $\widetilde q_R$.
Applying the reduction formula Eq.~(\ref{eq:antisym}) to the amplitude of the 
$\chi \chi \rightarrow g g$ diagram shown in 
Fig.~\ref{fig:fd}(d) allows us to write the amplitude for the diagram 
involving squark $\widetilde q_i$ as
\begin{equation}
  \mathcal{M}_i =\frac{- \sqrt 2 g_s^2 g^2_{r/l}}
	  {M_{\widetilde q_{i \, r/l}}^2}\int \frac{d^4 q}{(2 \pi)^4}
	  \text{Tr}\left[\mathcal{F}(q)_i^{\mu \nu, a b} \slash P 
	    \gamma_5 P_{R/L} \right]
	  \epsilon(k_1)^*_{\mu}\epsilon(k_2)^*_{\nu},
\end{equation}
where the neutralino-quark-squark couplings are defined for right- and 
left-handed squarks respectively as
\begin{equation}
  g_r = -\sqrt 2 N_{11} g' Q, \hspace{0.5in} 
  g_l = - \sqrt 2 N_{11} g' (T_3 - Q) + \sqrt 2 T_3 N_{12} g.
  \label{eq:grgl}
\end{equation}
Here $T_3$ is the squark isospin, $Q$ is the squark electric charge,
and $N_{11}$ and $N_{12}$ are the bino and wino components of the 
neutralino as defined in Ref. \cite{DarkSUSY}. 
We also define $P_{R/L}=(1\pm \gamma_5)/2$ in the usual 
way as the right- and left-handed projection operators.
The external gluon momenta are called $k_1$ and $k_2$, and $q$ is 
the momentum flowing in the loop.

The form factor $\mathcal{F}(q)_i^{\mu \nu, a b}$ for squark $\widetilde q_i$
and the corresponding internal quark $q_i$ is given explicitly by
\begin{eqnarray}
  \mathcal{F}(q)_i^{\mu \nu, a b} &=& 
  \frac{\slash q - \slash k_2 +m_{q_i}}{(q-k_2)^2-m_{q_i}^2} \gamma^{\nu}t^b 
  \frac{\slash q +m_{q_i}}{q^2-m_{q_i}^2} \gamma^{\mu} t^a 
  \frac{\slash q + \slash k_1 +m_{q_i}}{(q+k_1)^2-m_{q_i}^2}
  \nonumber \\
  & & + \frac{\slash q - \slash k_1 +m_{q_i}}{(q-k_1)^2-m_{q_i}^2}
  \gamma^{\mu}t^a \frac{\slash q +m_{q_i}}{q^2-m_{q_i}^2} \gamma^{\nu} 
  t^b \frac{\slash q + \slash k_2 +m_{q_i}}{(q+k_2)^2-m_{q_i}^2}.
\end{eqnarray}
The two terms in $\mathcal{F}(q)_i^{\mu \nu, a b}$ correspond to the 
two possible directions of fermion flow in Fig.~\ref{fig:fd}(d).
The diagrams with the neutralinos crossed are already included through
the use of the antisymmetrized spinors in Eq.~(\ref{eq:antisym}).

After summing over left and right squark states, 
the amplitude becomes
\begin{equation}
  \mathcal{M}_i=- \int \frac{d^4 q}{(2 \pi)^4} 2 \text{Tr}
  \left[\mathcal{F}(q)_i^{\mu \nu, a b} \gamma^{\alpha} 
    (V_i+A_i \gamma_5) \right] 
  P_{\alpha} \epsilon(k_1)^*_{\mu}\epsilon(k_2)^*_{\nu},
  \label{eq:Msummed}
\end{equation}
with 
\begin{equation}
  V_i = \frac{\sqrt 2 g_s^2}{4} \left( 
        \frac{g_r^2}{M_{\widetilde q_{i\, r}}^2} 
	- \frac{g_l^2}{M_{\widetilde q_{i\, l}}^2} \right) 
\hspace{0.5in}  
  A_i =  \frac{\sqrt 2 g_s^2}{4} \left( 
         \frac{g_r^2}{M_{\widetilde q_{i\, r}}^2} 
	 + \frac{g_l^2}{M_{\widetilde q_{i\, l}}^2} \right),
\end{equation}
with couplings $g_{r,l}$ given in Eq.~(\ref{eq:grgl}).

After integrating Eq.~(\ref{eq:Msummed}), the piece involving the vector 
coupling vanishes due to the conservation of vectorial current (CVC):
the contracted vector behaves as a divergence, so that the resulting 
vector coupling to multi-gluon states vanishes.  
Light quark masses can be neglected in the form factor if 
$m_b^2 \ll m_{\chi}^2$, and the top quark loop amplitude is suppressed
if $m_{\chi}^2 \ll m_t^2$.
In these limits, the loop amplitude sums over five massless quarks.  
Note that in the massless quark limit, the form factor 
$\mathcal{F}(q)_i^{\mu \nu, a b}$ becomes independent of the quark
flavor $i$.  The sum over quarks $q_i$ in the loop can then be 
factorized into a sum over the coupling factors $A_i$ times the universal
massless form factor $\mathcal{F}(q)_i^{\mu \nu, a b}$.

The gluon production amplitude contains a pseudovector triangle diagram.  
This can be transformed to a pseudoscalar triangle diagram via the axial 
anomaly, Eq.~(\ref{eq:anomaly}), where $\partial_{\mu}=-2i P_{\mu}$.  
The anomaly is computed by relating it to the decay of a fundamental 
pseudoscalar Higgs boson $A^0$ to two gluons via a heavy quark loop.
If the mass of the heavy quark $Q$ is sufficiently large, 
$m_{Q} \gg m_{A^0}$, then from Eqs.~(\ref{eq:anom2}) and (\ref{eq:anom3})
we have,
\begin{equation}
 2i P_{\alpha}  \sum A_i \bar q_i \gamma^{\alpha} \gamma_5 q_i
 = 2 m_{Q} \bar Q i \gamma_5 Q \sum A_i.
  \label{eq:chichiA}
\end{equation}
The amplitude becomes
\begin{eqnarray}
  \mathcal{M} &=& -2 m_{Q} \sum A_i \int \frac{d^4 q}{(2 \pi)^4} 
  \text{Tr}\left[\mathcal{F}(q)_{Q}^{\mu \nu, a b} \gamma_5\right]   
  \epsilon(k_1)^*_{\mu}\epsilon(k_2)^*_{\nu}
  \nonumber \\
  &=& -\frac{ m^2_{Q}}{\pi^2}  \sum A_i 
  C_0(0, 0, s, m^2_{Q}, m^2_{Q}, m^2_{Q}) 
  \text{Tr}\left[t^a t^b\right] \epsilon^{\mu \nu \alpha \beta} 
  k_{1 \alpha} k_{2 \beta}\epsilon(k_1)^*_{\mu}\epsilon(k_2)^*_{\nu},
\end{eqnarray}
where $s = 4 m_{\chi}^2$ and $C_0(0, 0, s, m^2_Q, m^2_Q, m^2_Q)$ is a
three-point Passarino-Veltman integral \cite{ref:passarino}.  In the limit of heavy quark mass, $m^2_{Q} \gg s$,
the three-point integral reduces to $-1/2 m_{Q}^2$.  
In this limit the dependence on the heavy quark mass drops out and
the amplitude becomes
\begin{equation}
  \mathcal{M} = \frac{1}{2} \delta^{ab} 
  \frac{\sum A_i}{2 \pi^2} \epsilon^{\mu \nu \alpha \beta} 
  k_{1 \alpha} k_{2 \beta} \epsilon(k_1)^*_{\mu}\epsilon(k_2)^*_{\nu},
\end{equation}
where we have used ${\rm Tr}[t^a t^b] = (1/2) \delta^{ab}$.
Squaring the amplitude and integrating over phase space gives the
leading-order ($\chi \chi \rightarrow g g$) annihilation cross section,
\begin{equation}
v_{\text{rel}}   \sigma_{LO}(\chi\chi \to gg) 
  = \frac{m_{\chi}^2}{64 \pi^5}\left( \sum_{q_i} A_i \right) ^2,
\label{eq:sigmavLO}
\end{equation}
where in our approximation the sum runs over the five light quarks $q_i$.
Our result agrees with that of, e.g., Ref.~\cite{Drees:1993bh} in
the limit $m_q = 0$.

\subsection{Beyond leading order}

As discussed in Sec.~\ref{sec:anomalyequation}, we can use the Adler-Bardeen
theorem \cite{ref:abj} and Eqs.~(\ref{eq:anom2}) and (\ref{eq:anom3})
to obtain the QCD corrections to the $\chi\chi \to gg$ annihilation 
cross section by exploiting the known results for pseudoscalar Higgs 
decays to gluon pairs, $A^0 \to gg$, beyond leading order.

The NLO  QCD corrections to the $A^0 \to gg$ partial width were 
calculated by Spira {\it et al}.~\cite{Spira:1995rr}.
In the heavy top quark limit, for which our anomaly relation is
valid, the NLO QCD corrections are given by a
multiplicative factor \cite{Spira:1995rr}, times the LO decay rate
\begin{equation}
  \Gamma_{NLO}(A^0 \rightarrow gg) = \Gamma_{LO}(A^0 \rightarrow gg)
  \left[1+\frac{\alpha_s}{\pi} \left( \frac{97}{4} - \frac{7}{6}N_f 
  + \frac{33 - 2 N_f}{6} \log \frac{\mu^2}{m_A^2}\right)
  \right],
  \label{eqn:spirawidth}
\end{equation} 
where $\mu$ is the renormalization scale.  The integer $N_f$ counts 
the number of quark flavors in the gluon splitting, with $N_f = 5$ 
for $m_b \ll m_{\chi} \ll m_t$.  
The diagrams that contribute to $A^0 \to gg$ at LO and NLO are shown
in Fig.~\ref{fig:Spiradiag}.  The NLO final states include $gg$, $ggg$,
and $g\bar q q$.

\begin{figure}[h]
\begin{center}
\includegraphics[scale=0.8]{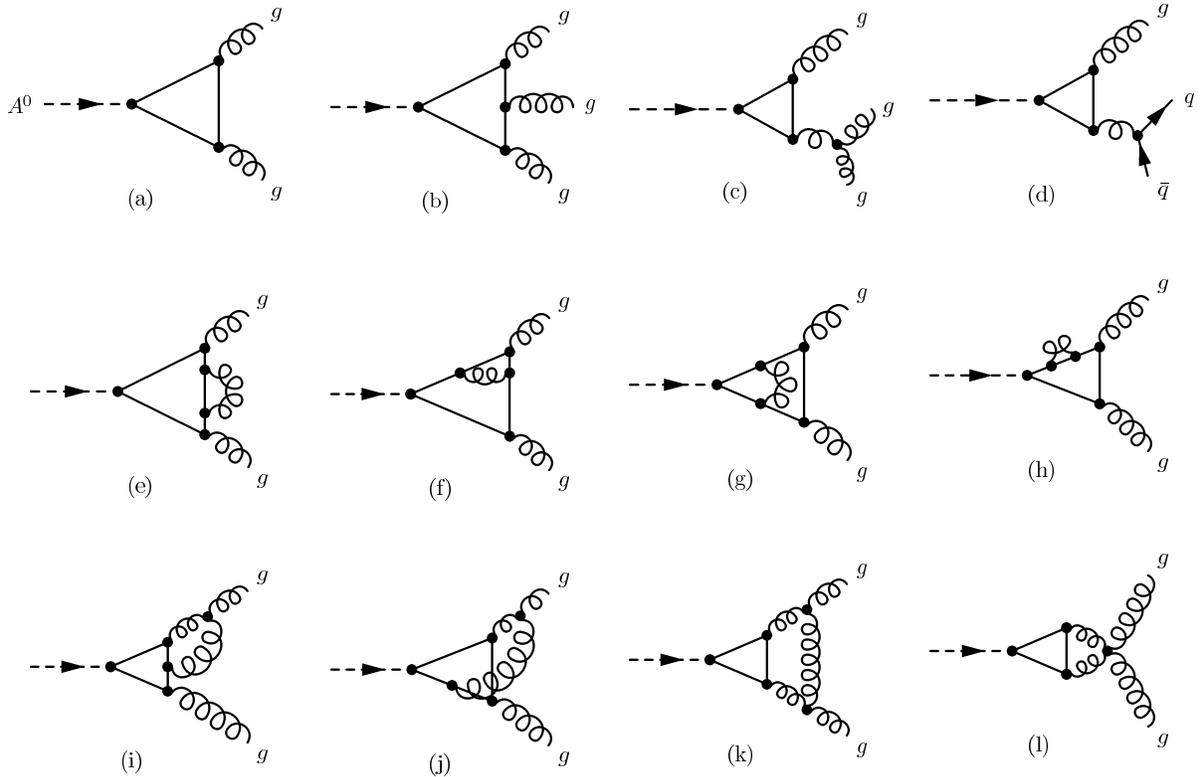}
\caption{LO and NLO diagrams for $A^0 \to gg$.
(a) is the leading order process, (b-d) are the real emission diagrams 
with three final-state particles, and (e-l) are virtual corrections 
to diagram (a).}
\label{fig:Spiradiag}
\end{center}
\end{figure}

\begin{figure}[h]
\begin{center}
\includegraphics[scale=1]{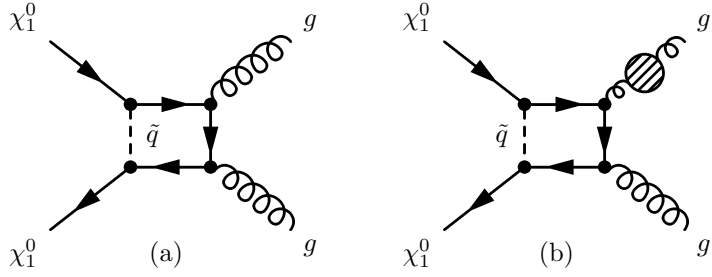}
\caption{$\chi\chi \to gg$ diagrams that supply a logarithmic factor to cancel
that from Fig.~\ref{fig:fd}.  The one-loop corrections to the gluon legs
include quark and gluon loops.}
\label{diaginterf}
\end{center}
\end{figure}

In $\chi\chi \to gg$ at NLO, a divergence occurs for the diagram in 
Fig.~\ref{fig:fd}(d) when the final-state
quarks are soft or collinear, in which case the gluon propagator diverges.  
This is the source of the logarithmic enhancement factor, $\log(m_{\chi}^2/m_q^2)$, found for this
diagram in Ref.~\cite{Flores:1989ru}.  However, this logarithmic term 
is precisely canceled by the renormalization of the strong coupling 
due to the quark bubble that appears in the virtual part of the NLO
correction, shown as the interference of the diagrams in
Fig.~\ref{diaginterf}.  This is the familiar cancellation of logarithmic
divergences guaranteed by the Kinoshita-Lee-Nauenberg 
theorem \cite{ref:kinoshita}.  A similar cancellation occurs for the
analogous diagrams in which the soft/collinear quarks are replaced with
gluons.

We now invoke the Adler-Bardeen theorem \cite{ref:abj} and
take over the NLO corrections to $A^0 \to gg$ to the $\chi\chi \to gg$ process in the zero-velocity limit.
In this correspondence, the pseudoscalar mass $m_A$ is replaced by the 
$\chi\chi$ center-of-mass energy, equal to $2 m_{\chi}$ in the zero-velocity
limit.  The NLO correction to the cross section for $\chi\chi \to gg$
follows immediately from Eqs.~(\ref{eq:sigmavLO}) and (\ref{eqn:spirawidth}),
\begin{eqnarray}
v_{\rm rel}  \sigma_{NLO}(\chi \chi \to gg) = 
  \frac{m_{\chi}^2}{64 \pi^5}\left( \sum A_i \right)^2 
  \left[1+\frac{\alpha_s}{\pi} \left( \frac{97}{4} - \frac{7}{6}N_f 
  + \frac{33 - 2 N_f}{6} \log \frac{\mu^2}{4 m_{\chi}^2}\right)
  \right]
\end{eqnarray}
We give explicitly the result for $\mu = 2 m_{\chi}$ and $N_f = 5$,
\begin{equation}
  v_{\rm rel} \sigma_{NLO}(\chi \chi \to gg) = \frac{m_{\chi}^2}{64 \pi^5}\left( \sum A_i \right)^2 
  (1 + 0.62),
    \label{eqn:spiraxv}
\end{equation} 
where we set $m_{\chi} = 100$ GeV.
The strong coupling is evaluated based on five-flavor running 
at the scale $\mu = 2 m_{\chi}$ where it appears both explicitly and
within the coefficient $A_i$.
We note that the above choice of $m_{\chi} = 100$ GeV is well above the 
current experimental limit \cite{ref:pdg}.

\section{Numerical results}
\label{sec:results}

In this section we examine the validity of our assumption of massless
quarks in the loop for the LO $\chi \chi \to gg$ calculation, show the
improvement in the renormalization scale dependence obtained in going to
NLO, and compare the $\chi \chi \to gg$ annihilation cross section
to that of the leading-order tree level process $\chi \chi \to q \bar q$
through $t$-channel squark exchange.

We have assumed in our use of the anomaly equation that the five light
quarks running in the loop for $\chi \chi \to gg$ were massless, and we
neglected the heavy top quark contribution.
In Fig. \ref{fig:xvanom}, we test this assumption for the LO cross 
section by comparing our approximation to the full cross section including
the quark mass dependence (we continue to use the $1/\widetilde M^2$ 
approximation for the squark propagator).  
We plot the full cross section normalized
to our five-massless-quark approximation as a function of $m_{\chi}$.
The full formula differs from our approximation by less than 10\% for
6~GeV~$< m_{\chi} < 110$~GeV.  For heavier neutralinos, the top quark
loop starts to have a significant effect, with destructive interference
occurring between the top loop and the lighter quark loops.  Two of the 
top quarks in the loop go on shell at $m_{\chi} = m_t$, leading to the 
large dip in the cross section.  For neutralinos lighter than about 
6~GeV, the nonzero mass of the bottom quark begins to play a significant
role, leading to the dip at lower masses.

\begin{figure}[!h]
\begin{center}
\includegraphics[scale=0.3,angle=-90]{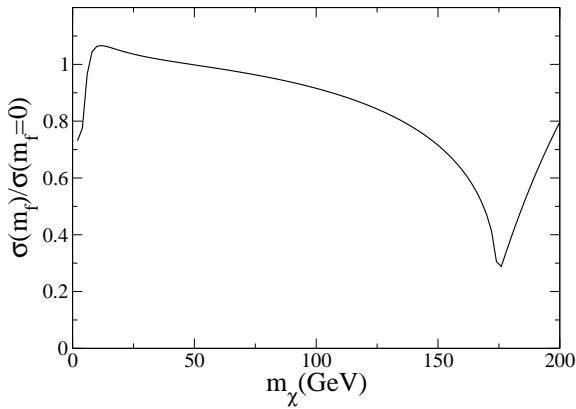}
\caption{The effect of quark masses on 
$v_{\text rel} \sigma_{LO}(\chi \chi \to g g) $.
Shown is the full cross section including the quark mass
dependence, normalized to the approximate cross section obtained by
setting $m_q = 0$ for the five light quarks and $m_t = \infty$,
for a common squark mass $\widetilde M=200$ GeV and a pure bino
neutralino, $N_{11} = 1$, $N_{1j} = 0$ ($j \neq 1$).  
}
\label{fig:xvanom}
\end{center}
\end{figure}

In an all-orders calculation, physical observables cannot depend on the
renormalization scale $\mu$.  The $\mu$ dependence of our predictions
is an artifact of computing to a finite order in perturbation theory.
The $\mu$ dependence can then be used to estimate the size of the 
uncomputed higher-order corrections.
The dependence of the $\chi \chi \to g g$ annihilation cross section
on the renormalization scale at LO and NLO is shown 
in Fig.~\ref{fig:mu}(a) for $\widetilde M = 200$ GeV, 
$m_{\chi} = 100$~GeV, and a pure
bino neutralino, $N_{11} = 1$, $N_{1j} = 0$ ($j \neq 1$).\footnote{We note 
here that the renormalization scale dependence of the
tree-level $\chi\chi \to q \bar q$ cross section arises only from the
running quark mass in the $s$-wave contribution \cite{Drees:1993bh}.  
Since the cross section
for this process is dominated by the quark-mass-independent $p$-wave part
during freeze-out in the early universe, the scale dependence is negligible 
at tree-level.}

Varying $\mu$ by a factor of two in either direction from 
the central value $\mu = 2 m_{\chi}$ yields a scale dependence of 
$\pm 16\%$ at LO and $\pm 9\%$ at NLO.
The corresponding annihilation cross sections including scale uncertainty 
are shown in Fig.~\ref{fig:mu}(b) as a function of $m_{\chi}$.

\begin{figure}[htbp]
\begin{center}
\includegraphics[scale=0.28, angle=-90]{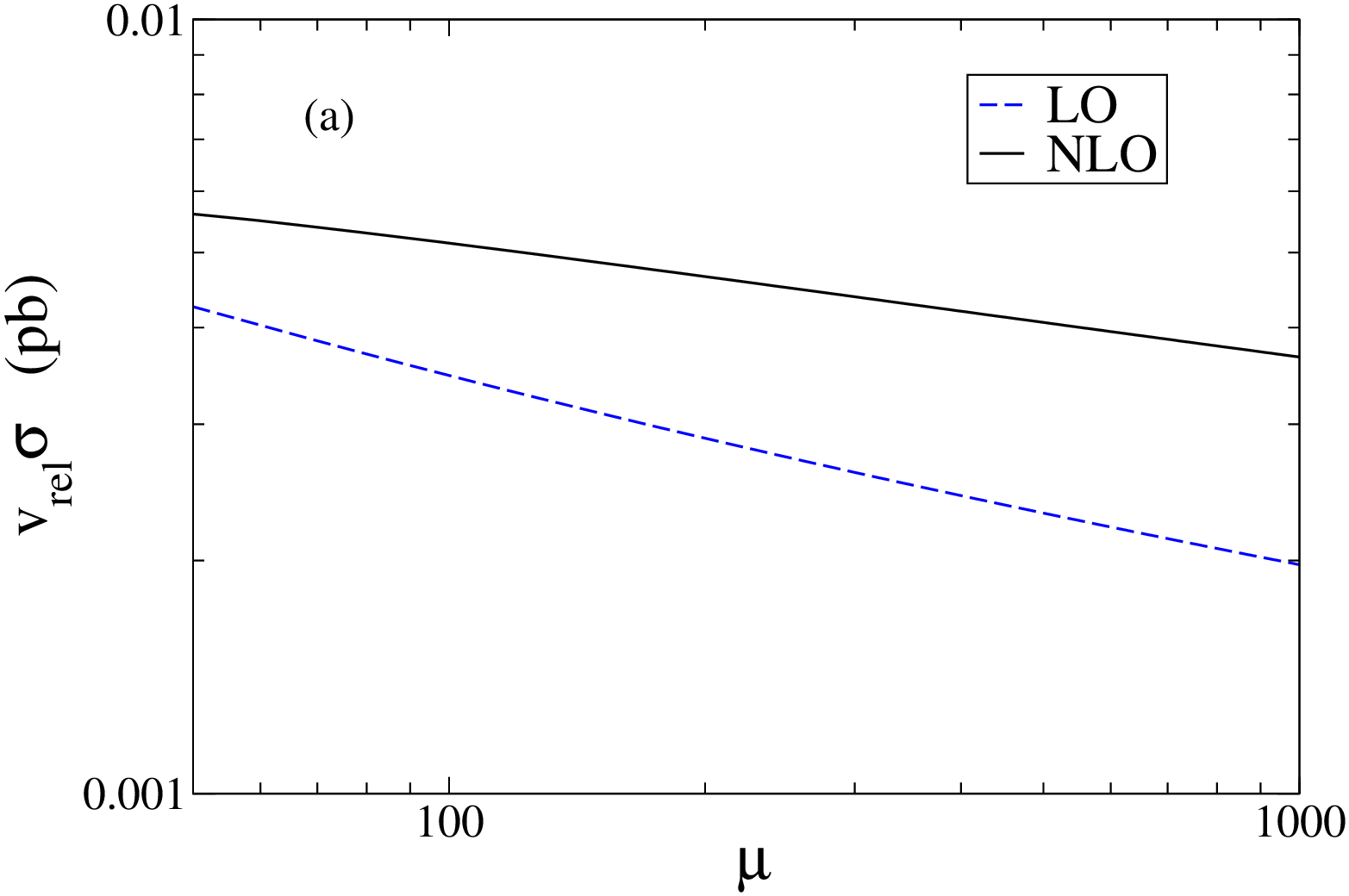}
\includegraphics[scale=0.28, angle=-90]{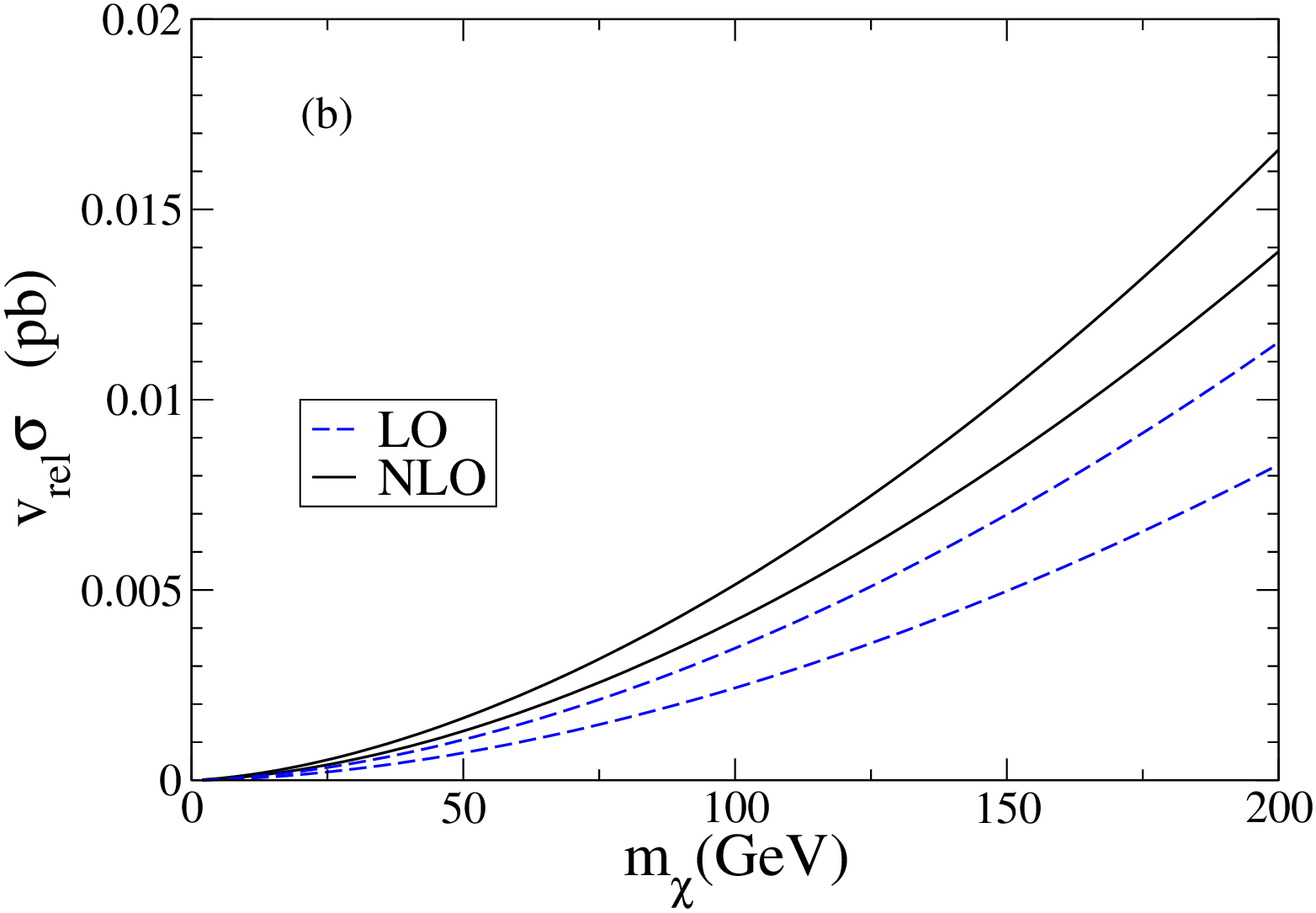}
\caption{Cross section for $\chi\chi \to gg$ at LO and NLO for a 
pure bino neutralino with $\widetilde M = 200$ GeV.  
(a) Dependence on the renormalization scale
$\mu$ for $m_{\chi} = 100$ GeV.  (b) Dependence on $m_{\chi}$ showing the
renormalization scale dependence in the band $\mu = m_{\chi}$, $4 m_{\chi}$.}
\label{fig:mu}
\end{center}
\end{figure}

Exact cross section formulae for all tree-level two-to-two neutralino 
annihilation processes are given in Ref.~\cite{Nihei:2002ij}.  Using 
the expansion of the thermally averaged 
cross section in terms of $x = T/m_{\chi}$,
\begin{equation}
 \langle v_{\text rel}  \sigma  \rangle = a + b x,
\end{equation}
one can compare the leading tree-level process, $\chi\chi \to q \bar q$ via
$t$-channel squark exchange, to our results for $\chi \chi \to gg$ 
at NLO given in Eq.~(\ref{eqn:spiraxv}) both for annihilation during
freeze-out in the early universe, $x \sim 1/20$, and for annihilation 
in the galactic halo today, $x \sim 0$ (corresponding to $v/c \sim 10^{-3}$).
The cross sections for the two annihilation processes are shown in 
Fig.~\ref{fig:xvN12}(a) as a function of $m_{\chi}$ for 
$\widetilde M = 200$ GeV and a pure bino neutralino, 
with $x = 1/20$ and $x = 0$.  For $\chi \chi \to q \bar q$ we sum
the cross section over the five light final-state quark flavors and 
neglect $\chi\chi$ annihilation into lepton pairs.
We use the running quark mass $m_q(\mu)$, which serves to resum the
leading logarithmic QCD corrections to this process from soft gluon 
radiation \cite{Drees:1993bh}, and take the renormalization scale 
$\mu = 2 m_{\chi}$.
The annihilation cross section for $\chi\chi \to gg$ is dominated
by the $s$-channel component, and so we show only one curve for
$x=1/20$ and $x=0$.
The tree-level cross section for $\chi\chi \to q \bar q$ depends strongly
on the relative velocity of the neutralinos, because the $s$-channel 
cross section is suppressed by the final-state quark mass.  Thus
the annihilation cross section at $x=0$, which comes only from the $s$-wave 
component, is quite small and is comparable to that from $\chi\chi \to gg$. 
At $x=1/20$, on the other hand, the $\chi\chi \to q \bar q$ cross section 
is dominated by the $p$-wave component and is larger than $\chi\chi \to gg$
by almost a factor of 100.

In Fig.~\ref{fig:xvN12}(b) we show the corresponding K-factors,
defined as the ratio of the total cross section, $\chi \chi \to q \bar q + gg$,
to the tree-level $\chi\chi \to q \bar q$ cross section.  This gives a 
measure of the relative importance of the $\chi \chi \to gg$ component
of the total annihilation cross section.  We see that during freeze-out, 
$x \sim 1/20$, the 
$\chi\chi \to gg$ contribution is quite small and $K = 1.01-1.02$ over the 
range of $m_{\chi}$ considered.  In the present epoch, however, with 
$x \sim 0$, the K-factor is considerably larger, $K = 1.3-30$ depending
on the mass of the neutralino.  Such a large K-factor will impact 
annihilation branching fractions today, changing the gamma ray flux 
from the galactic halo and the neutrino flux from inside the Sun 
\cite{Drees:1993bh,dmrev}.
We note also that because both the 
$\chi\chi \to q \bar q$ and $\chi\chi \to gg$ annihilation cross sections 
have the same leading $1/\widetilde M^4$ dependence on the squark
mass, these K-factors will not depend significantly on the common squark mass
scale.

\begin{figure}[!h]
\begin{center}
\includegraphics[scale=0.28,angle=-90]{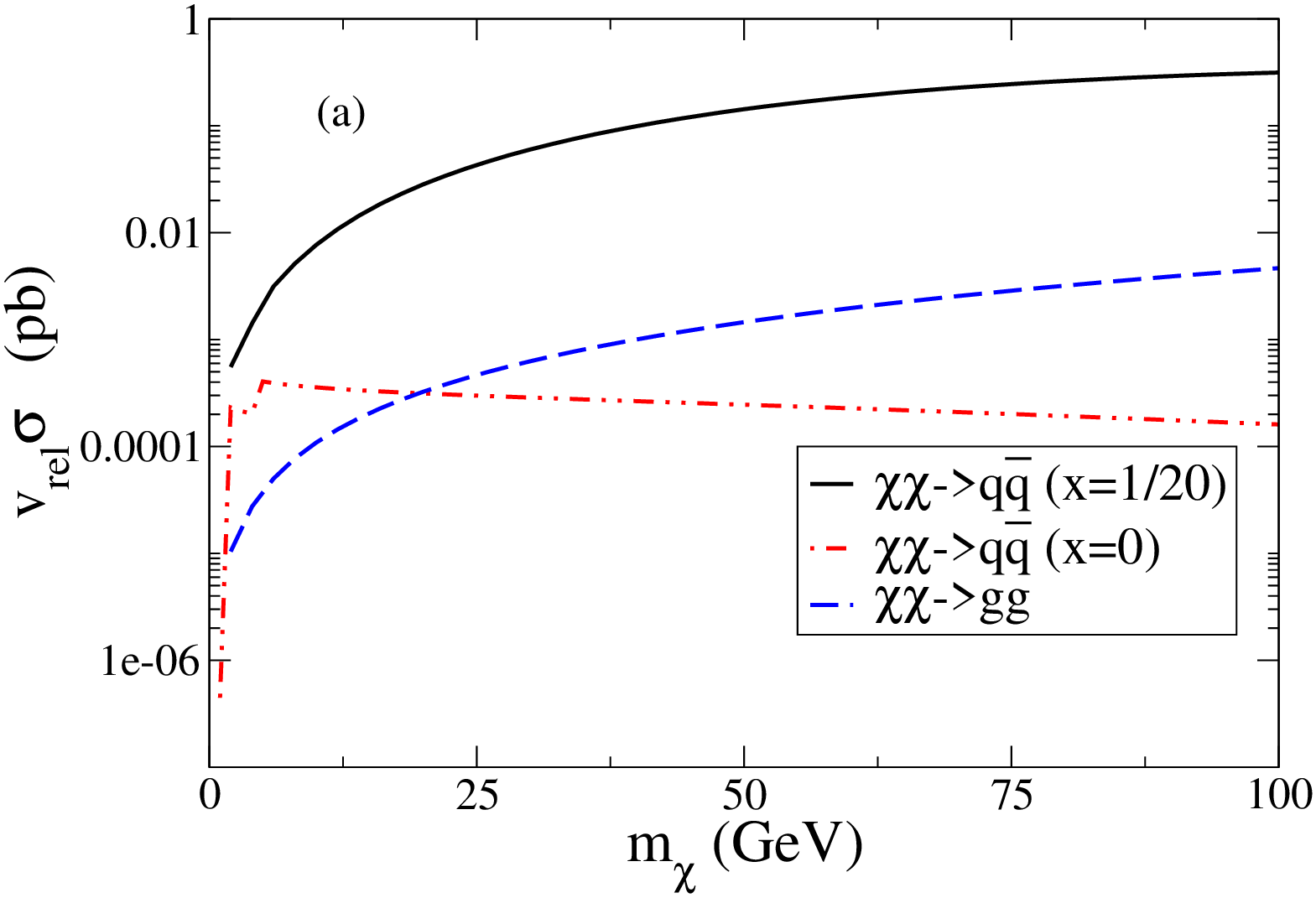} 
\includegraphics[scale=0.28, angle=-90]{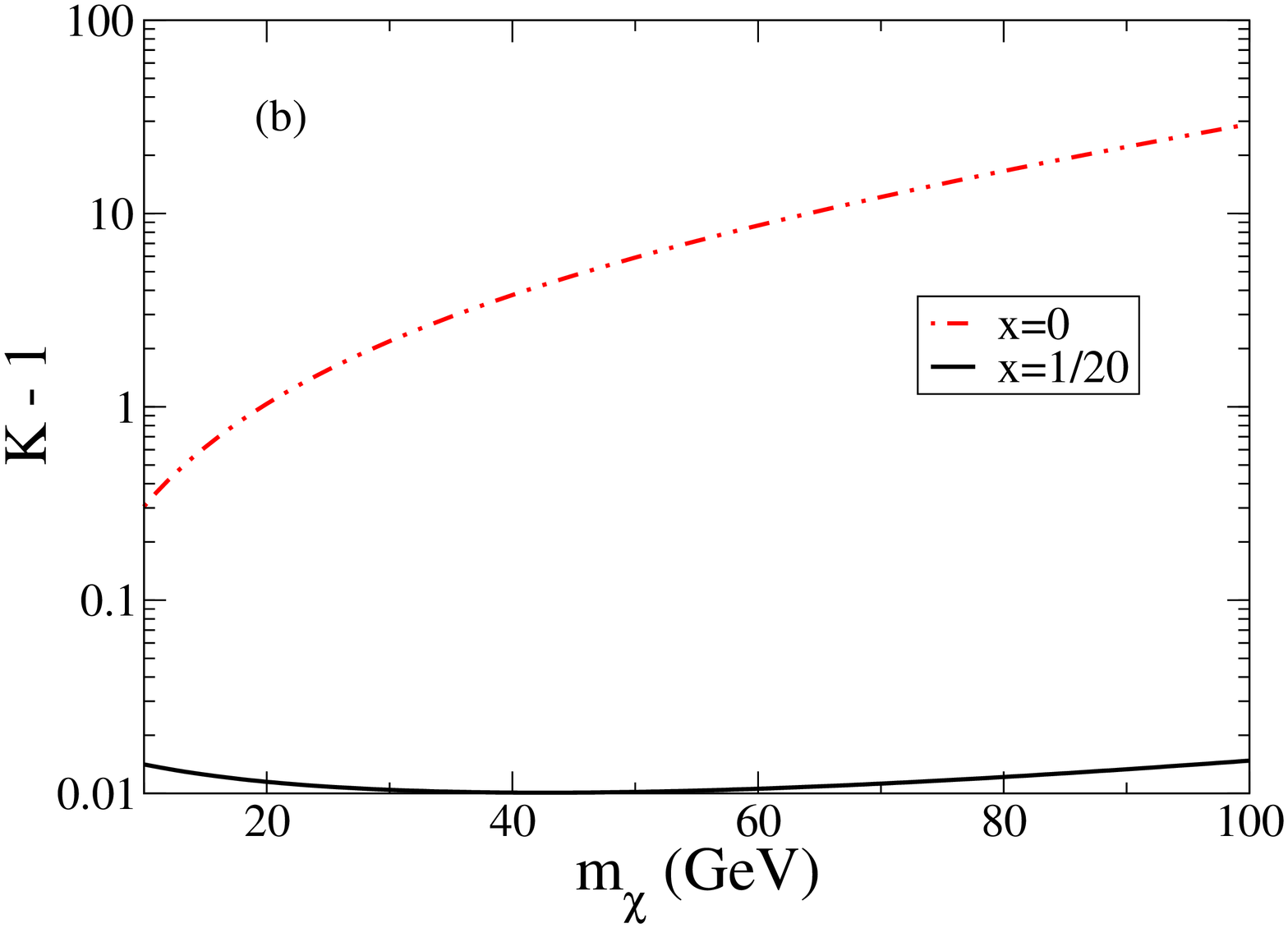} 
\caption{(a) Annihilation cross sections for $\chi \chi \to q \bar q$
through $t$-channel squark exchange [diagram~\ref{fig:fd}(a)] 
and $\chi \chi \to gg$ at NLO for $\widetilde M = 200$ GeV
and a pure bino neutralino.  We show $\chi \chi \to q \bar q$ for 
both $x=1/20$, corresponding to freeze-out in the early universe,
and $x=0$, corresponding to annihilation in the galactic halo at the
present time.  The steps in the $\chi\chi \to q \bar q$ cross section
for $x=0$ at low $m_{\chi}$ are due to quark mass thresholds.
(b) Corresponding K-factors.  We plot $K-1$, which is the ratio of 
annihilation cross sections of $\chi\chi \to gg$ to $\chi\chi \to q \bar q$,
for $x=0$ and $x=1/20$ as shown in (a).}
\label{fig:xvN12}
\end{center}
\end{figure}

We now exhibit the dependence of the annihilation cross section on the
neutralino composition.  Because we have worked in the zero-quark-mass
limit in our calculation of $\chi\chi \to gg$, we have neglected the 
couplings of Higgsinos to the internal quark loop, which are proportional
to the quark mass.  We thus consider only mixed bino-wino neutralinos.
We can then parameterize the mixing coefficients $N_{1j}$ in terms 
of a bino-wino mixing angle $\phi$ as
\begin{equation}
  N_{11} = \cos \phi, \qquad \qquad
  N_{12} = \sin \phi, \qquad \qquad
  N_{13} = N_{14} = 0.
\end{equation}
In Fig.~\ref{fig:xvmix} we again compare the leading tree-level process, 
$\chi\chi \to q \bar q$ via $t$-channel squark exchange, to our results 
for $\chi \chi \to gg$ at NLO.  The cross sections for the two 
annihilation processes are shown in Fig.~\ref{fig:xvmix}(a) as a 
function of the bino-wino mixing angle $\phi$ for $m_{\chi} = 50$~GeV,
and $\widetilde M = 200$~GeV, with $x=1/20$ and $x=0$.  
Again we take the renormalization
scale $\mu = 2 m_{\chi}$.  There is a large
enhancement of both annihilation cross sections if the neutralino has a 
large wino component, $\phi \sim 90^{\circ}$, due to the stronger coupling
of the wino to a quark-squark pair.
In Fig.~\ref{fig:xvmix}(b) we show the corresponding K-factors.  The 
nontrivial dependence of the K-factors on $\phi$ arises from the
interference among the five light quark loop diagrams that contribute
to $\chi\chi \to gg$.
By contrast, there is no 
interference between the amplitudes for $\chi\chi \to q \bar q$ with
different flavor squarks in the $t$-channel because the internal squark
flavor is fixed by the external quark flavor.

\begin{figure}[!h]
\begin{center}
\includegraphics[scale=0.28,angle=-90]{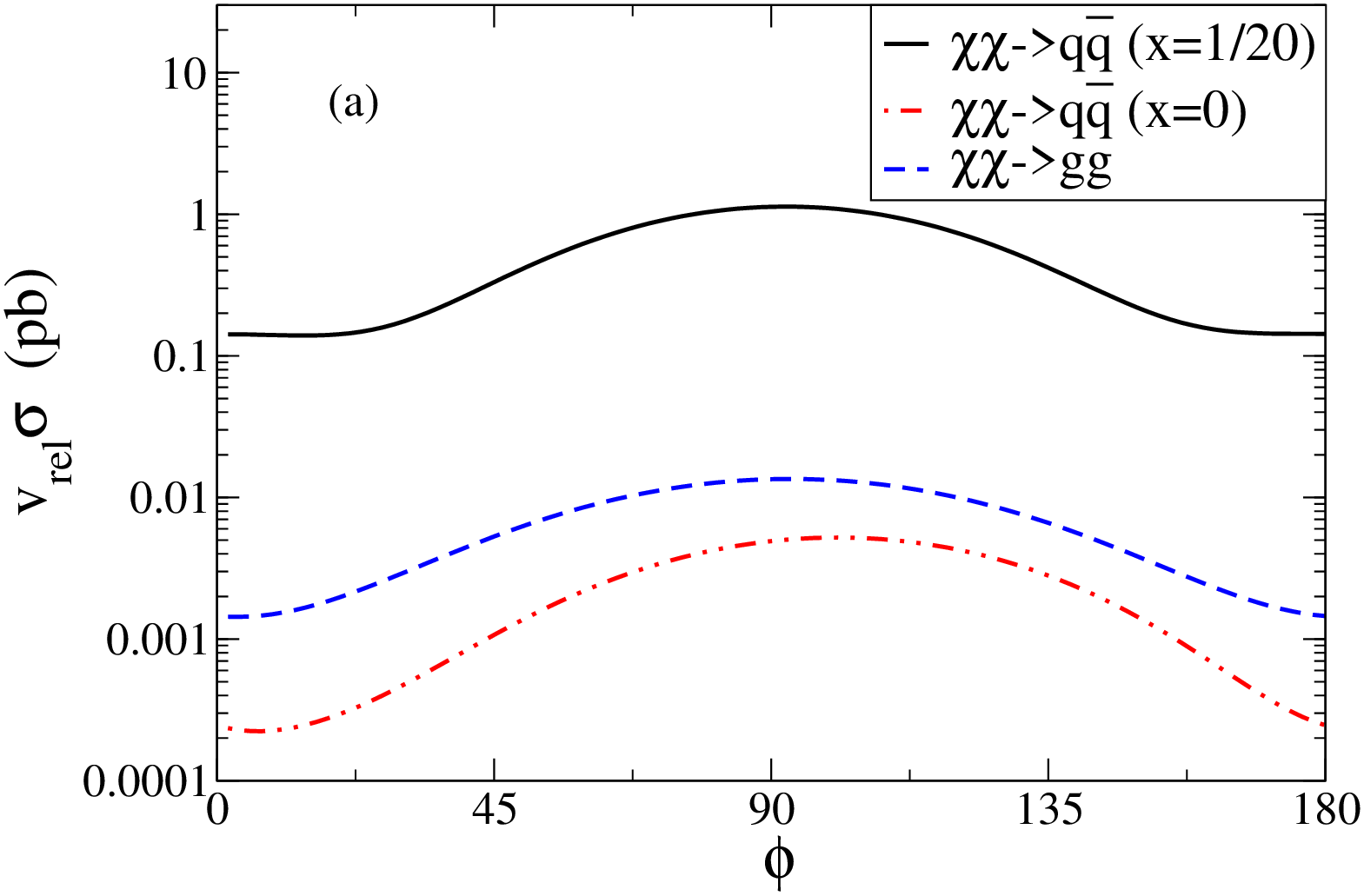}
\includegraphics[scale=0.28,angle=-90]{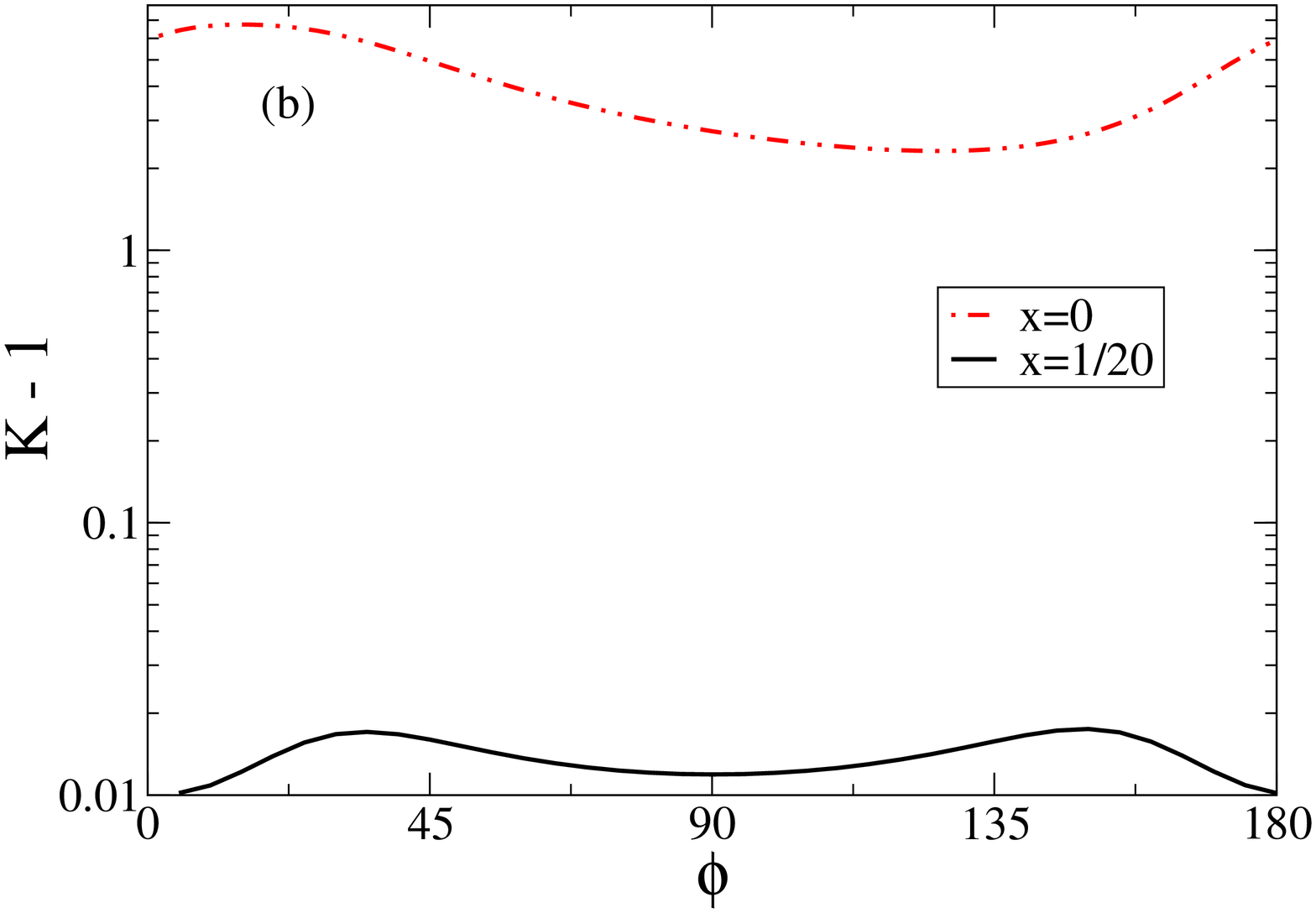}
\caption{(a) Annihilation cross sections for $\chi \chi \to q \bar q$
and $\chi \chi \to gg$ as in Fig.~\ref{fig:xvN12} as a function of the 
bino-wino mixing angle $\phi$, for $m_{\chi} = 50$ GeV.
Pure bino corresponds to $\phi = 0^{\circ}, 180^{\circ}$ and pure wino 
to $\phi = 90^{\circ}$.
(b) Corresponding K-factors.}
\label{fig:xvmix}
\end{center}
\end{figure}

Finally, we note that the next-to-next-to-leading order (NNLO)
corrections to $A^0 \to gg$ have been computed in
Ref.~\cite{Chetyrkin:1998mw}.  One may be tempted to take this correction
over to $\chi\chi \to gg$ in the same way as the NLO correction.
However, at NNLO the $A^0 \to gg$ decay receives a contribution from the
interference between the $G^{(a)}_{\mu\nu} \widetilde G^{(a) \mu \nu}$
operator and a $\partial_{\mu} \bar q \gamma^{\mu} \gamma_5 q$ operator
generated at two-loop level.  Once effective operators other than
$G^{(a)}_{\mu\nu} \widetilde G^{(a) \mu \nu}$ appear, our use of the anomaly
equation no longer applies.  A proper treatment of $\chi\chi \to gg$ at
NNLO would thus require a new calculation of the operator matching
conditions and renormalization.

%-------------------------------------------------------------------
\section{Conclusions}
\label{sec:conclusions}

We reviewed the dependence of the main neutralino annihilation processes
on various suppression factors -- the Higgsino fraction, the quark and
squark masses, and the relative neutralino velocity -- and identified
the dominant $\chi \chi \to q \bar q$ annihilation process for a gaugino-like
neutralino as due to a dimension-six operator in the zero-velocity limit.
This dimension-six operator contains the divergence of the axial vector
current of the quarks $q \bar q$, which leads to the well-known quark 
mass suppression of the annihilation cross section.  This quark mass
suppression can be lifted in two ways: either through corrections to the
dimension-six operator involving the anomalous triangle diagram, or
by going to dimension-eight.  We focused on the anomalous triangle diagram,
which describes neutralino annihilation to gluon pairs.  In the approximation
of massless quarks running in the loop, the anomaly equation relates
$\chi\chi \to gg$ to the seemingly unrelated process of pseudoscalar
decay to gluon pairs via a very heavy quark loop.  We used this 
relation to compute $\chi\chi \to gg$ in terms of the decay process
$A^0 \to gg$.  Further, taking advantage of the Adler-Bardeen theorem
which guarantees that the anomaly equation is valid to all orders in 
$\alpha_s$, we extracted the NLO QCD corrections to $\chi \chi \to gg$
from the known corresponding results for $A^0 \to gg$ and wrote them
as a simple multiplicative factor that can be easily inserted into
numerical neutralino annihilation codes.  For 
$m_{\chi} = 100$ GeV and $\mu=2 m_{\chi}$, the NLO QCD corrections 
increase the annihilation cross section by 62\%.
The NLO corrections also reduce the residual renormalization scale 
dependence of the $\chi\chi \to gg$ annihilation cross section 
from $\pm 16\%$ to $\pm 9\%$.

Our NLO results were computed in the approximation 
$m_b \ll m_{\chi} \ll m_t$.  This approximation yields a LO $\chi\chi \to gg$
cross section within 10\% of the exact result for 
6~GeV~$< m_{\chi} < 110$~GeV.
We finally compared our results for $\chi\chi \to gg$ at NLO to the 
dominant $\chi \chi \to q \bar q$ cross section in this neutralino
mass range.  For neutralino annihilation during freeze-out in the 
early universe, our results for $\chi\chi \to gg$ at NLO constitute 
only $1-2\%$ of the dominant cross section for a gaugino-like neutralino
and are thus of little importance for computing the relic neutralino abundance.
However, for neutralino annihilation at the present time 
the relative neutralino
velocity is much lower, leading to a much smaller tree-level $\chi\chi\to
q \bar q$ cross section.  In this situation, the $\chi\chi \to gg$ cross
section can be as large or larger than $\chi\chi \to q \bar q$, so that
NLO corrections can have a significant impact on the computation of
gamma ray and neutrino fluxes from neutralino annihilation in the 
galatic halo and inside the Sun, respectively.

%--------------------------------------------------------------
\begin{acknowledgments}
This work was supported in part by the U.S.~Department of Energy
under grants DE-FG02-95ER40896 and DE-FG05-84ER40173
and in part by the Wisconsin Alumni Research Foundation.  
We thank W.~Bardeen and M.~Drees for helpful comments.
VB thanks the Aspen Center for Physics for hospitality 
during the completion of this work.  
\end{acknowledgments}

%--------------------------------------------------------------


\begin{thebibliography}{9}

\bibitem{dmrev}
 G.~Bertone, D.~Hooper and J.~Silk,
  %``Particle dark matter: Evidence, candidates and constraints,''
  Phys.\ Rept.\  {\bf 405}, 279 (2005)
  [arXiv:hep-ph/0404175].
  %%CITATION = HEP-PH 0404175;%%
  
\bibitem{Bennett:2003bz}
C.~L.~Bennett {\it et al.}  [WMAP Collaboration],
  %``First Year Wilkinson Microwave Anisotropy Probe (WMAP) Observations:
  %Preliminary Maps and Basic Results,''
  Astrophys.\ J.\ Suppl.\  {\bf 148}, 1 (2003)
  [arXiv:astro-ph/0302207];
  %%CITATION = ASTRO-PH 0302207;%%
D.~N.~Spergel {\it et al.}  [WMAP Collaboration],
  %``First Year Wilkinson Microwave Anisotropy Probe (WMAP)
  %Observations:
  %Determination of Cosmological Parameters,''
  Astrophys.\ J.\ Suppl.\  {\bf 148}, 175 (2003)
  [arXiv:astro-ph/0302209];
  %%CITATION = ASTRO-PH 0302209;%%
M.~Tegmark {\it et al.}  [SDSS Collaboration],
  %``Cosmological parameters from SDSS and WMAP,''
  Phys.\ Rev.\ D {\bf 69}, 103501 (2004)
  [arXiv:astro-ph/0310723].
  %%CITATION = ASTRO-PH 0310723;%%

\bibitem{Tauber}
  J.~A.~Tauber,
  ``The PLANCK Mission,''
in {\sl IAU Symposium}, Manchester, UK, 15-18 August 2000, 
ed. M. Harwit, p. 493.

\bibitem{ref:sps}
 S.~Abel {\it et al.}  [SUGRA Working Group Collaboration],
  %``Report of the SUGRA working group for run II of the Tevatron,''
  arXiv:hep-ph/0003154.
  %%CITATION = HEP-PH 0003154;%%

\bibitem{Djouadi:2001kb}
  A.~Djouadi, M.~Drees, P.~Fileviez Perez and M.~Muhlleitner,
  %``Loop induced Higgs and Z boson couplings to neutralinos and  implications
  %for collider and dark matter searches,''
  Phys.\ Rev.\ D {\bf 65}, 075016 (2002)
  [arXiv:hep-ph/0109283].
  %%CITATION = HEP-PH 0109283;%%

\bibitem{Yang:1950rg}
  C.~N.~Yang,
  %``Selection Rules For The Dematerialization Of A Particle 
  %Into Two Photons,''
  Phys.\ Rev.\  {\bf 77}, 242 (1950).
  %%CITATION = PHRVA,77,242;%%

\bibitem{Rudaz:1989ij}
S.~Rudaz,
  %``On The Annihilation Of Heavy Neutral Fermion Pairs Into Monochromatic
  %Gamma-Rays And Its Astrophysical Implications,''
  Phys.\ Rev.\ D {\bf 39}, 3549 (1989);
  %%CITATION = PHRVA,D39,3549;%%
L.~Bergstrom,
  %``Radiative Processes In Dark Matter Photino Annihilation,''
  Phys.\ Lett.\ B {\bf 225}, 372 (1989).
  %%CITATION = PHLTA,B225,372;%%

\bibitem{ref:adlera}
S.~L.~Adler,
``Perturbation Theory Anomalies,''
in {\it Lectures on Elementary Particles and Quantum Field Theory},
Vol. 1, ed. S.~Deser, M.~Grisaru and H.~Pendleton
(M.I.T. Press, Cambridge, MA), pp. 3--164 (1970).

\bibitem{ref:abj}
S.~L.~Adler and W.~A.~Bardeen,
  %``Absence Of Higher Order Corrections In The Anomalous Axial Vector
  %Divergence Equation,''
  Phys.\ Rev.\  {\bf 182}, 1517 (1969).
  %%CITATION = PHRVA,182,1517;%%

\bibitem{Spira:1995rr}
M.~Spira, A.~Djouadi, D.~Graudenz and P.~M.~Zerwas,
  %``Higgs boson production at the LHC,''
  Nucl.\ Phys.\ B {\bf 453}, 17 (1995)
  [arXiv:hep-ph/9504378].
  %%CITATION = HEP-PH 9504378;%%

\bibitem{Drees:1993bh}
  M.~Drees, G.~Jungman, M.~Kamionkowski and M.~M.~Nojiri,
  %``Neutralino annihilation into gluons,''
  Phys.\ Rev.\ D {\bf 49}, 636 (1994)
  [arXiv:hep-ph/9306325].
  %%CITATION = HEP-PH 9306325;%%

\bibitem{Flores:1989ru}
R.~Flores, K.~A.~Olive and S.~Rudaz,
  %``Radiative Processes In Lsp Annihilation,''
  Phys.\ Lett.\ B {\bf 232}, 377 (1989).
  %%CITATION = PHLTA,B232,377;%%

\bibitem{Bergstrom:1988}
L.~Bergstrom and H.~Snellman,
  %``Observable Monochromatic Photons From Cosmic Photino Annihilation,''
  Phys.\ Rev.\ D {\bf 37}, 3737 (1988).
  %%CITATION = PHRVA,D37,3737;%%

\bibitem{Kuhn:1979}
  J.~H.~Kuhn, J.~Kaplan and E.~G.~O.~Safiani,
  %``Electromagnetic Annihilation Of E+ E- Into Quarkonium States With Even
  %Charge Conjugation,''
  Nucl.\ Phys.\ B {\bf 157}, 125 (1979).
  %%CITATION = NUPHA,B157,125;%%
  
\bibitem{DarkSUSY}
 P.~Gondolo, J.~Edsjo, P.~Ullio, L.~Bergstrom, M.~Schelke and E.~A.~Baltz,
  %``DarkSUSY: Computing supersymmetric dark matter properties numerically,''
  JCAP {\bf 0407}, 008 (2004)
  [arXiv:astro-ph/0406204].
  %%CITATION = ASTRO-PH 0406204;%%

\bibitem{ref:passarino}
  G.~Passarino and M.~J.~G.~Veltman,
  %``One Loop Corrections For E+ E- Annihilation Into Mu+ Mu- In The Weinberg
  %Model,''
  Nucl.\ Phys.\ B {\bf 160}, 151 (1979).
  %%CITATION = NUPHA,B160,151;%%

\bibitem{ref:kinoshita}
T.~Kinoshita,
  %``Mass Singularities Of Feynman Amplitudes,''
  J.\ Math.\ Phys.\  {\bf 3}, 650 (1962);
  %%CITATION = JMAPA,3,650;%%
T.~D.~Lee and M.~Nauenberg,
  %``Degenerate Systems And Mass Singularities,''
  Phys.\ Rev.\  {\bf 133}, B1549 (1964).
  %%CITATION = PHRVA,133,B1549;%%

\bibitem{ref:pdg}
S.~Eidelman {\it et al.}  [Particle Data Group],
  %``Review of particle physics,''
  Phys.\ Lett.\ B {\bf 592}, 1 (2004).
  %%CITATION = PHLTA,B592,1;%%

\bibitem{Nihei:2002ij}
  T.~Nihei, L.~Roszkowski and R.~Ruiz de Austri,
  %``Exact cross sections for the neutralino WIMP pair-annihilation,''
  JHEP {\bf 0203}, 031 (2002)
  [arXiv:hep-ph/0202009].
  %%CITATION = HEP-PH 0202009;%%

\bibitem{Chetyrkin:1998mw}
K.~G.~Chetyrkin, B.~A.~Kniehl, M.~Steinhauser and W.~A.~Bardeen,
  %``Effective {QCD} interactions of CP-odd Higgs bosons at three
  %loops,''
  Nucl.\ Phys.\ B {\bf 535}, 3 (1998)
  [arXiv:hep-ph/9807241].
  %%CITATION = HEP-PH 9807241;%%

  
\end{thebibliography}
\end{document}